\documentclass{elsart}
\usepackage{graphicx}
\usepackage{amssymb}
\begin{document}
\begin{frontmatter}
% Title, authors and addresses
% use the thanksref command within \title, \author or \address for footnotes;
% use the corauthref command within \author for corresponding author footnotes;
% use the ead command for the email address,
% and the form \ead[url] for the home page:
%\title{Title\thanksref{label1}}
% \thanks[label1]{}
% \author{Name\corauthref{cor1}\thanksref{label2}}
% \ead{email address}
% \ead[url]{home page}
% \thanks[label2]{}
% \corauth[cor1]{}
% \address{Address\thanksref{label3}}
% \thanks[label3]{}
\title{Wealth Dynamics on Complex Networks}
% use optional labels to link authors explicitly to addresses:
% \author[label1,label2]{}
% \address[label1]{}
% \address[label2]{}
\author[1,2]{Diego Garlaschelli}
\and
\author[2,3]{Maria I. Loffredo}
\address[1]{Dipartimento di Fisica, Universit\`a di Siena, Via Roma 56, 53100 Siena ITALY}
\address[2]{INFM UdR Siena, Via Roma 56, 53100 Siena ITALY}
\address[3]{Dipartimento di Scienze Matematiche ed Informatiche, Universit\`a di Siena, Pian dei Mantellini 44, 53100 Siena ITALY}
\begin{abstract}
We study a model of wealth dynamics [Bouchaud and M\'ezard 2000, \emph{Physica A} \textbf{282}, 536] which mimics transactions among economic agents. The outcomes of the model are shown to depend strongly on the topological properties of the underlying transaction network. The extreme cases of a fully connected and a fully disconnected network yield power-law and log-normal forms of the wealth distribution respectively. We perform numerical simulations in order to test the model on more complex network topologies. We show that the mixed form of most empirical distributions (displaying a non-smooth transition from a log-normal to a power-law form) can be traced back to a heterogeneous topology with varying link density, which on the other hand is a recently observed property of real networks.
\end{abstract}
\date{23 January 2004}
\end{frontmatter}
% main text

\section{Introduction}
How networks self-organize into complex structures is one of the main interests of modern statistical mechanics \cite{strogatz,albert}. A large amount of data has been assembled which suggests that many different biological, social and technological networks display non-trivial properties, strongly deviating from what expected in simple random graphs \cite{albert}. This has stimulated the proposal of theoretical models aimed at reproducing the observed topological features by means of basic mechanisms, invoking either growth processes \cite{strogatz,albert} or `hidden' underlying principles \cite{fitness}.

Understanding the topological properties of real-world networks is fundamental also because they affect the outcomes of dynamical processes defined on them:
strong quantitative and qualitative effects of network topology have been highlighted in models describing synchronization \cite{strogatz}, spread of epidemics \cite{epidemics} and wealth exchange \cite{bouchaud,soumaSW,wehia}.
In the present paper we focus on the latter topic by discussing how the form of the statistical distribution of the wealth of a set of economic agents depends on the topological properties of the transaction network defined among them. Our analyses are performed using the model of wealth exchange proposed by Bouchaud and M\'ezard \cite{bouchaud}. This problem is interesting for various reasons: firstly, the outcomes of the model can be directly compared with empirical data \cite{wehia,pareto,gibrat,levy,souma}; secondly, wealth distributions range in general between two typical forms which interestingly correspond to the two extreme cases of a fully connected and a fully disconnected network; finally, the problem is a first step towards the more general one of understanding how network topology affects the outcomes of a random process governed by stochastic differential equations.

\section{The Bouchaud-M\'ezard Model}
In the model of Bouchaud and M\'ezard \cite{bouchaud} ($BM$ in the following) the wealth $w_i$ of an agent $i$ ($i=1,\dots N$) is assumed to be governed by the following set of stochastic differential equations:
\begin{equation}
\label{BM}
\dot{w}_i(t)=\eta_i(t)w_i(t)+\sum_{j\ne i}J_{ij}w_j(t)-\sum_{j\ne i}J_{ji}w_i(t)
\end{equation}
where the $\eta_i(t)$'s are independent Gaussian variables of mean $m$ and variance $2\sigma^2$ (accounting for the intrinsic variation of wealth) and $J_{ij}=Ja_{ij}/N$ is the fraction of agent $j$'s wealth flowing into agent $i$'s wealth, which is directly related to the adjacency matrix $a_{ij}$ of the underlying transaction network (which in the following we assume to be undirected: $a_{ij}=a_{ji}=1$ if $i$ and $j$ exchange wealth and $a_{ij}=a_{ji}=0$ if they do not). Equation (\ref{BM}) is interpreted in the Stratonovich sense.

The above model can be easily studied within the Fokker-Planck approach in the mean-field limit of a fully connected network ($a_{ij}=1\quad\forall i\ne j$) \cite{bouchaud}. In such a case the form of the equilibrium (long-term) wealth distribution $\rho(x)$ (expressed as a function of the normalized wealth $x_i\equiv w_i/\sum_j w_j$) can be obtained analytically \cite{bouchaud} and reads
\begin{equation}
\label{eq}
\rho(x)\propto\frac{\exp[(1-\beta_{mf})/x]}{x^{1+\beta_{mf}}}
\end{equation}
where $\beta_{mf}\equiv 1+J/\sigma^2$. The tail (large $w$ limit) of this distribution clearly follows a power-law behaviour with exponent $-1-\beta_{mf}$. 

In the opposite case of a network with no link ($J_{ij}=0\quad\forall i,j$), the sums appearing in eq.(\ref{BM}) vanish and the process is purely multiplicative. The Central Limit Theorem then states that for large $N$ the distribution of $log(x)$ is normal, or in other words the distribution of $x$ is log-normal. A useful family of distributions is therefore given by 
\begin{equation}
\label{both}
\rho(x)\propto\frac{\exp(-\alpha/x)}{x^{1+\beta}}
\end{equation}
which is almost undistinguishable from a log-normal one when $\beta\approx 0$ and displays power-law tails of the form $x^{-1-\beta}$ for $\beta>0$. In particular, the case $\alpha=\beta-1=J/\sigma^2$ corresponds to the mean-field prediction (\ref{eq}).

Real wealth distributions \cite{pareto,levy} always display power-law tails with exponent $-1-\beta$ where $\beta$ is usually near to, but often smaller than, one. By contrast, the mean-field limit of the $BM$ model always predicts that $\beta>1$.  Another property of real data is in general a non-smooth transition from a left log-normal-like part of the distribution and a right power-law tail \cite{wehia,pareto,gibrat,levy,souma}. Both properties are not reproduced by the simple mean-field case, and in the following we ask if a suitable, and possibly more realistic, choice of the transaction network yields the desired outcomes.

\section{Random graphs, regular lattices and scale-free networks}
The first point of having an exponent $\beta<1$ out of the $BM$ model was already explored in \cite{bouchaud}, where the authors showed that when the network is a random graph \cite{albert} (characterized by a Poissonian degree distribution) $\beta$ is systematically smaller than the mean-field prediction $\beta_{mf}$. Clearly, when the connection probability $p\approx 0$ all vertices are isolated and the log-normal-like distribution is recovered, while $p\to 1$ yields the mean-field case. Therefore in the random graph case the distribution is well described by eq. (\ref{both}), where $\beta\to 0$ for $p\to 0$ and $\beta\to\beta_{mf}$ for $p\to 1$.

\begin{figure}[h]
\begin{center}
\includegraphics[width=.49\textwidth]{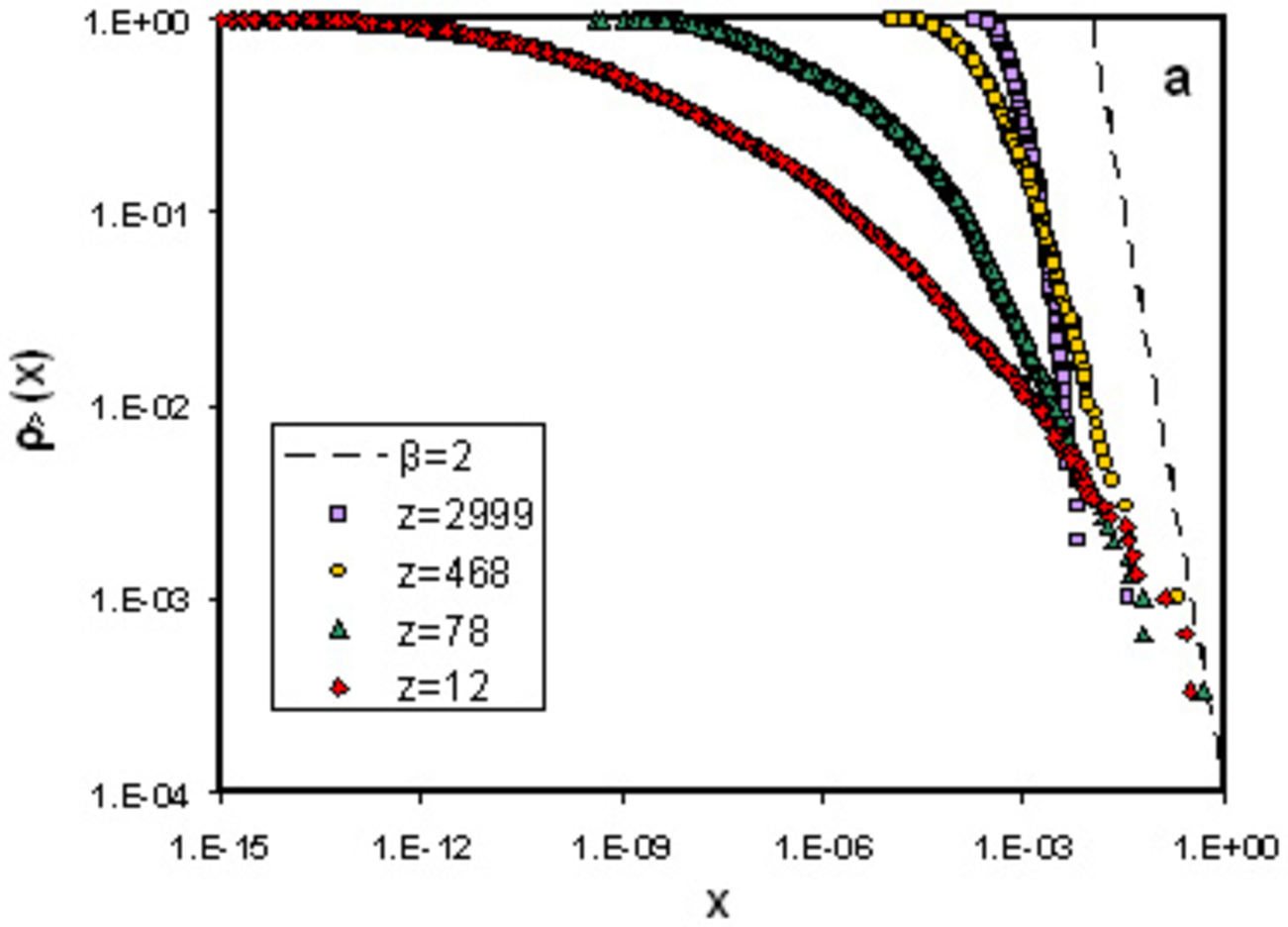}
\includegraphics[width=.49\textwidth]{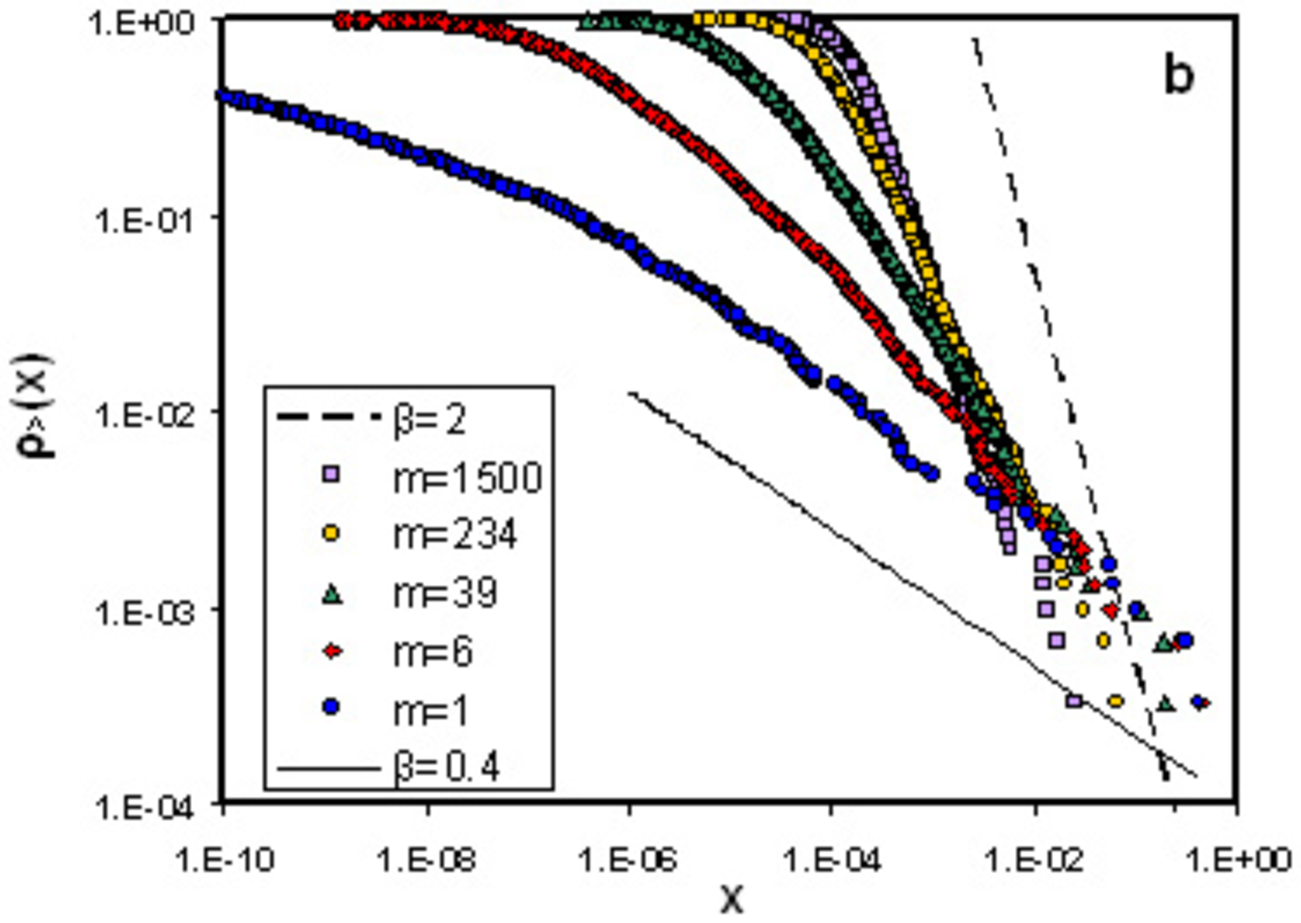}
\end{center}
\caption[]{Cumulative wealth distributions generated by the $BM$ model on networks with $N=3000$ vertices. The dashed line always corresponds to the slope predicted by the mean-field theory ($\beta=1+J/\sigma^2=2$). a) Regular ring for various choices of the degree $z$. b) $BA$ scale free network for various choices of the number $m$ of vertices injected at each timestep. The solid line is drawn as a reference for the range of $\beta$.}
\label{unif}
\end{figure}

We now show that a similar result holds when the network is a regular ring where each vertex is connected to its $z/2$ closest neighbours (that is, each vertex has degree $z$ and the degree distribution equals $P(k)=\delta_{kz}$). When $z=2$ one has a simple linear structure, and the log-normal-like ($\beta\approx 0$) behaviour is expected \cite{bouchaud}. By contrast, when $z\to N-1$ the network approaches the fully connected topology and one expects $\beta\to\beta_{mf}$. As we show in fig.\ref{unif}a, this is indeed the case and the cumulative wealth distribution $\rho_>(x)\equiv\int_x^\infty\rho(x')dx'$ approaches the mean-field shape $\rho_>(x)\propto x^{-\beta_{mf}}$ as $z$ increases (for simplicity, here and in the following we always set $J/\sigma^2=1$ so that $\beta_{mf}=2$). Our results are in accordance with those of ref. \cite{soumaSW}, where the authors also introduce a rewiring mechanism to obtain small-world networks \cite{albert} for various values of $z$ in order to have a wealth distribution displaying a right power-law tail and a left log-normal part. They find that for a suitable choice of the parameters this indeed occurs, however the two parts of the distribution are always connected in a smooth fashion, in constrast with the non-smooth behaviour of real data.
%%%%%%%%%%%%%%%% Fig1: ring e BA %%%%%%%%%%%%%%%%%  
%%%%%%%%%%%%%%%%%%%%%%%%%%%%%%%%%%%%%%%%%%%%%%%%%% 

We now turn to scale-free networks generated by the Barab\'asi-Albert ($BA$) mechanism \cite{albert}, where at each timestep one introduces a new vertex and $m$ new links connected to preexisting vertices chosen with probability proportional to their degree. The resulting degree distribution has the form $P(k)\propto k^{-3}$ \cite{albert}. We first generated a network with a given value of $m$ and then let the $BM$ model run on it. As shown in fig. \ref{unif}b, the cumulative distribution is always of the power-law type, with exponent $\beta$ ranging from the quite small value $0.4$ to $\beta=\beta_{mf}=2$. 

The results discussed so far show that the wealth distribution generated by the $BM$ model is strongly affected by the topology of the underlying network. The effect of the topology of the networks considered so far is however only quantitative, since essentially it consists in determining the exponent of the distribution (\ref{both}). This allows to obtain values of the exponent smaller than the mean-field prediction, which was one of the necessary points we discussed at the beginning. However, in order to generate a \emph{mixed} wealth distribution with the aforementioned non-smooth transition from a log-normal-like body to a power-law tail, deep qualitative effects of the topology are required, and more complicated networks must be considered.

\section{Heterogeneously linked networks}
\begin{figure}[h]
\begin{center}
\includegraphics[width=.49\textwidth]{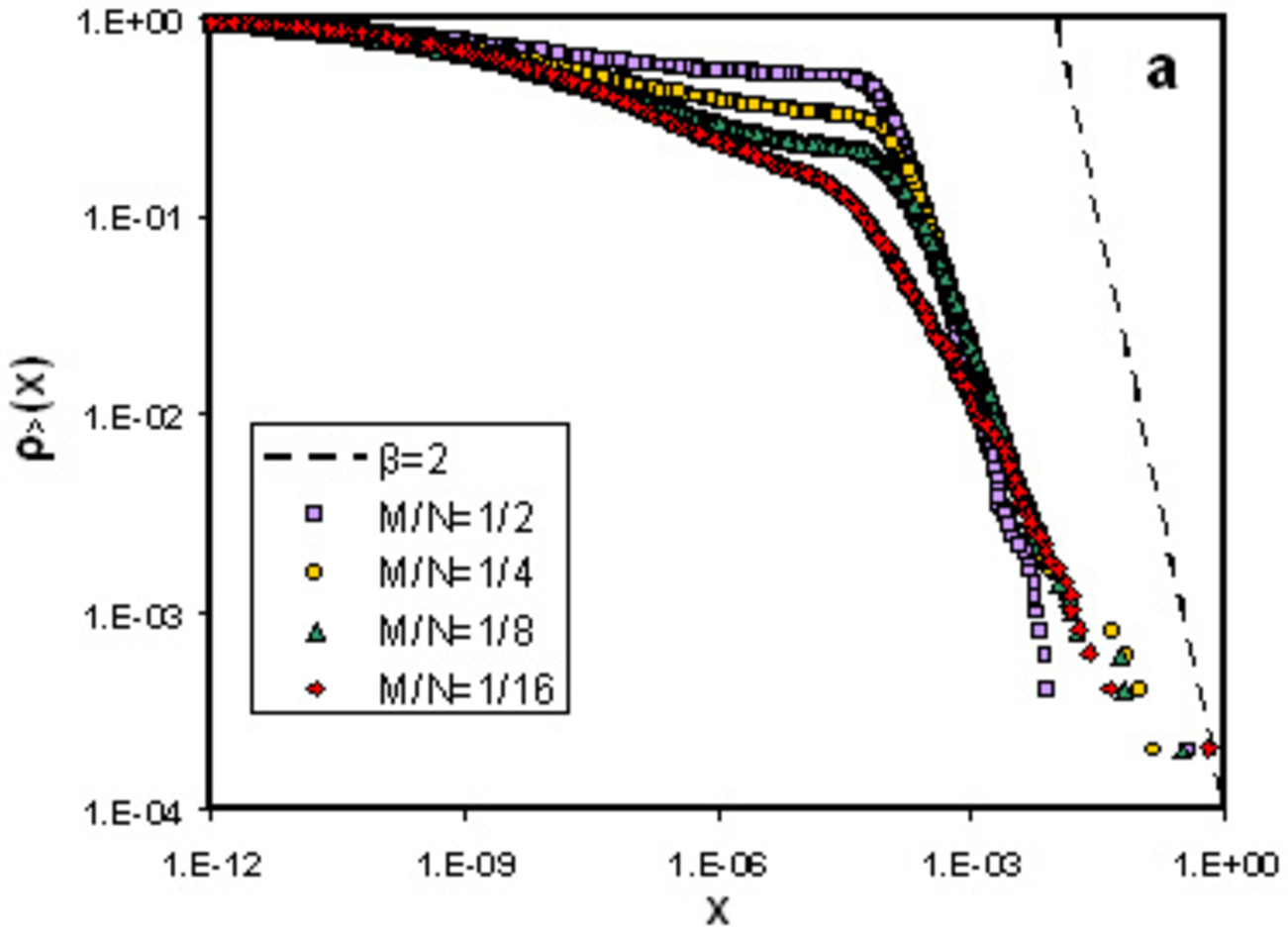}
\includegraphics[width=.49\textwidth]{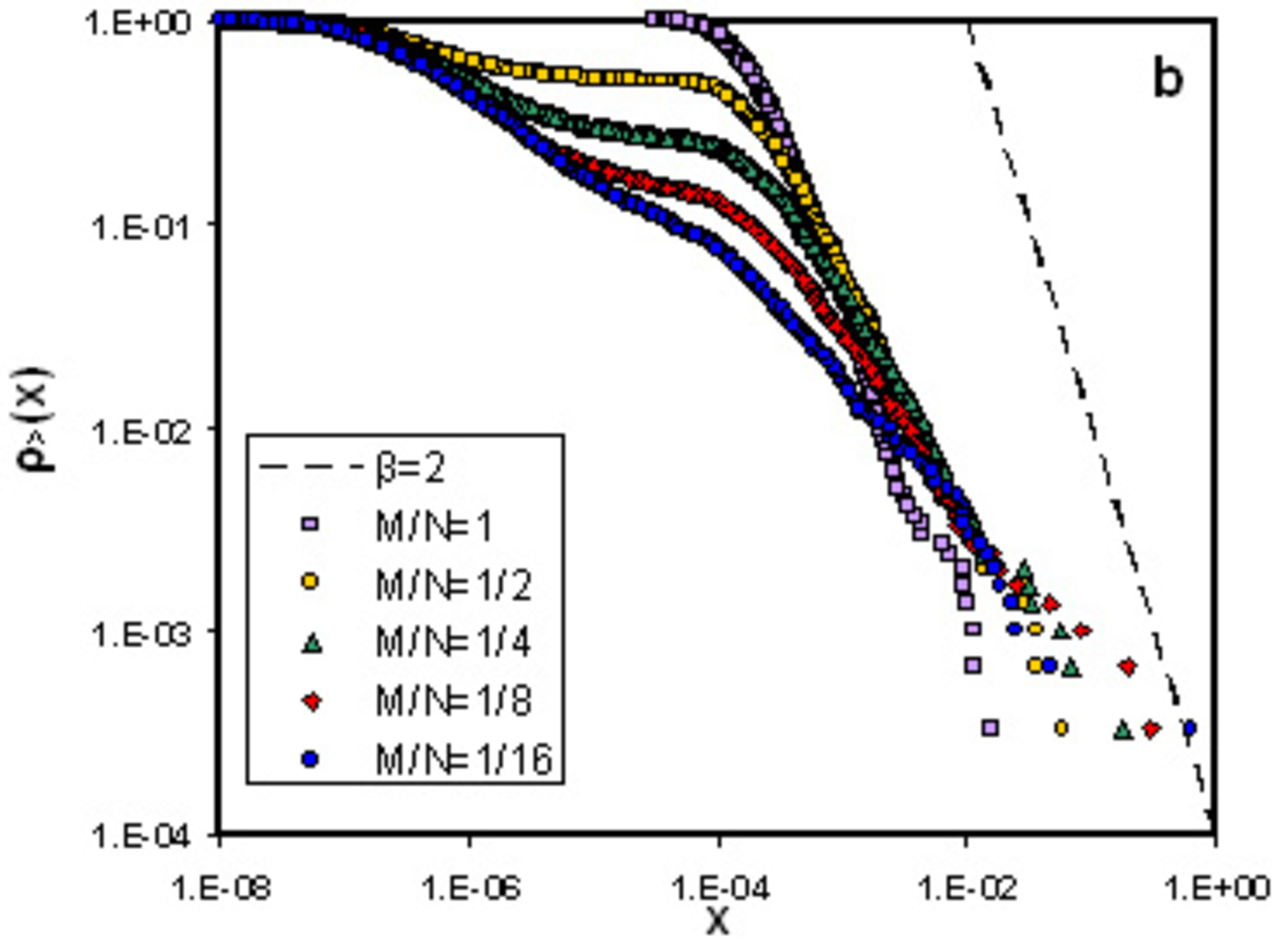}
\end{center}
\caption[]{Cumulative wealth distributions generated by the $BM$ model on a) mixed networks and b) octopus networks for various choices of the ratio $M/N$ (dashed line: mean-field prediction $\beta=2$). The number of vertices is $N=5000$.}
\label{mixed}
\end{figure}

We now consider a very simple example to show how the mixed form can arise. Consider a network consisting of a fully connected sub-network with $M$ vertices, plus a set of $N-M$ completely isolated vertices. Clearly, in this case we expect the wealth distribution to be of the form $\rho(x)=(M/N)\rho_{pl}(x)+(1-M/N)\rho_{ln}(x)$, where $\rho_{pl}$ and $\rho_{ln}$ indicate the power-law and log-normal predictions respectively. Then, by tuning the ratio $M/N$ it is possible to vary the relative weights of the two contributions. The results are reported in fig. \ref{mixed}a, showing that the mixed form is indeed observed.

We can now introduce more complex models characterized by the same basic ingredient of a dense `core' and a sparse `periphery'. In the absence of empirical data on real transaction networks, all choices are quite arbitrary and here we simply report the results obtained for an \emph{octopus} network having a random graph with $M$ vertices as the core and a set of $N-M$ `tentacles' with single edges linked to randomly chosen vertices in the core. The results are shown in fig. \ref{mixed}b, and in a wide intermediate range the form of the distribution is indeed very similar to the observed one \cite{wehia,pareto,gibrat,levy,souma}.

%%%%%%%%%%%%%%%% Fig2: mixed e octopus %%%%%%%%%%%%%%%%%  
%%%%%%%%%%%%%%%%%%%%%%%%%%%%%%%%%%%%%%%%%%%%%%%%%% 

\section{Discussion}
We showed that a crucial ingredient in order to have a realistic wealth distribution out of the $BM$ model is a heterogeneous link density in the network.
Detecting regions with high link density is crucial in social networks, where clustered portions of the network usually indicate communities of tightly interacting people. More in general, the way the network can display heterogeneously linked regions depends not simply on the degree distribution $P(k)$, but on higher-order properties such as the correlations between the degree of adjacent vertices \cite{newman} and the hierarchical dependence of the local clustering properties on the degree \cite{hierarchy}. Indeed, in the present paper we showed that $P(k)$ alone does not determine the desired properties of the wealth distribution. In order to relate the empirical form of wealth distributions to higher-order nontrivial topological properties would require the knowledge of the corresponding transaction networks. While this is a very hard task at the individual level, it is possible for the trade network of world countries, where the vertices are countries and links represent import/export trade relationships. Indeed, a recent study \cite{wtw} of the topological properties of the World Trade Web ($WTW$) revealed the presence of degree correlations and of a hierarchical organization of the network. Interesting, these are exactly the properties that, according to our analysis, are expected to determine the peculiar shape of the corresponding distribution of the `wealth' of vertices, which is this case is the Gross Domestic Product ($GDP$) of countries. Indeed, the $GDP$ distribution of world countries closely resembles the mixed form shown in fig. \ref{mixed}b \cite{wehia}. Future studies on the interplay between network topology of the $WTW$ and the $GDP$ dynamics, currently in preparation, will clarify this aspect in a more quantitative way. 

% The Appendices part is started with the command \appendix;
% appendix sections are then done as normal sections
% \appendix
% \section{}
% \label{}

\end{document}